\documentclass[aip,rsi,reprint,graphicx,amsmath,amssymb,amsfonts,floatfix]{revtex4-1} 
\usepackage{subfigure}
\usepackage{psfrag}
\usepackage[dvips]{graphicx}
\usepackage{color}
\usepackage{aas_macros}

\newcommand{\degs}{^\circ}

\begin{document}

\title{Modulation of CMB polarization with a warm rapidly-rotating half-wave plate on the Atacama B-Mode Search (ABS) instrument} %Title of paper

\author{A.~Kusaka}
\affiliation{Department of Physics, Princeton University, Princeton, New Jersey, 08544 USA}

\author{T.~Essinger-Hileman}
\email[]{essinger@pha.jhu.edu}
\author{J.~W.~Appel}
\affiliation{Department of Physics, Princeton University, Princeton, New Jersey, 08544 USA}
\affiliation{Department of Physics and Astronomy, The Johns Hopkins University, Baltimore, Maryland, 21218 USA}

\author{P.~Gallardo}
\affiliation{Department of Physics, Cornell University, Ithaca, New York, 14853 USA}

\author{K.~D.~Irwin}

\affiliation{National Institute of Standards and Technology, 325 Broadway MC 817.03, Boulder, Colorado, 80305 USA}
\affiliation{Department of Physics, Stanford University, Stanford, California 94305 USA}

\author{N.~Jarosik}
\affiliation{Department of Physics, Princeton University, Princeton, New Jersey, 08544 USA}

\author{M.~R.~Nolta}
\affiliation{The Canadian Institute for Theoretical Astrophysics, University of Toronto, Toronto, Ontario, M5S 3H8, Canada}

\author{L.~A.~Page}
\author{L.~P.~Parker}
\affiliation{Department of Physics, Princeton University, Princeton, New Jersey, 08544 USA}

\author{S.~Raghunathan}
\affiliation{Department of Astronomy, Universidad de Chile, Santiago, Chile}

\author{J.~L.~Sievers}
\affiliation{Department of Physics, Princeton University, Princeton, New Jersey, 08544 USA}
\affiliation{Astrophysics and Cosmology Research Unit, University of KwaZulu-Natal, Durban, 4041, South Africa}

\author{S.~M.~Simon}
\author{S.~T.~Staggs}
\author{K.~Visnjic}
\affiliation{Department of Physics, Princeton University, Princeton, New Jersey, 08544 USA}

\date{\today}

\begin{abstract}
We evaluate the modulation of Cosmic Microwave Background
 (CMB) polarization using a rapidly-rotating, half-wave plate (HWP) on
 the Atacama B-Mode Search (ABS). After demodulating
 the time-ordered-data (TOD), we find a significant
 reduction of atmospheric fluctuations. The demodulated TOD is stable on time scales of 500--1000\,seconds, corresponding to frequencies of 1--2\,mHz.
 This facilitates recovery of cosmological information at large
 angular scales, which are typically available only from balloon-borne or
 satellite experiments.
 This technique also
 achieves a sensitive measurement of celestial polarization
 without differencing the TOD of paired
 detectors sensitive to two orthogonal linear polarizations.
This is the first demonstration of the ability to remove atmospheric contamination at these levels from a ground-based platform using a rapidly-rotating HWP.

\end{abstract}


\maketitle

\section{Introduction}
Measurements of the Cosmic Microwave Background (CMB) temperature
anisotropies provide a particularly clean probe of the universe at the
time of decoupling, $\sim 400,000$ years after the Big Bang, and have
allowed constraints to be placed on cosmological parameters at the 1\%
 level. \cite{2013ApJS..208...19H, 2013arXiv1303.5076P,
 2013arXiv1301.0824S, 2011ApJ...743...28K} Measurement of CMB
 polarization at large angular scales also provides information about
 the very early universe and physics at grand-unified-theory energy
 scales through its sensitivity to a primordial gravitational-wave
 background (GWB). Many models of inflation predict that such a GWB should exist, and a GWB would leave a unique odd-parity pattern, termed ``B modes,'' in the CMB polarization.\cite{1997PhRvL..78.2054S,1997PhRvL..78.2058K}
 
The magnitude of the gravitational-wave B-mode signal is known to be less
than 100\,nK.~\cite{2010ApJ...711.1123C,2011ApJ...741..111Q,2013ApJS..208...19H,2013arXiv1303.5076P} From the ground, this signal must be viewed through an
atmosphere of 5--20\,K in antenna temperature,
which can fluctuate by tens of mK on minute time
scales.~\cite{2000ApJ...543..787L} Rapid signal modulation is a well-known technique first used by Dicke in radio astronomy,~\cite{1946RScI...17..268D} and CMB polarization experiments have used a number of modulation schemes.~\cite{2003ApJS..145..413J, 2004ApJ...610..625F, 2005ApJS..159....1B, 2012ApJ...760..145Q, 2010A&A...520A...4B, 2006PhDT........33S, 2003PhRvD..68d2002O, 2005ApJ...624...10L, CBI_instrument_2002, 2009ApJ...694.1664C, 2013ApJ...765...64M} Rapid modulation of linear polarization by a half-wave plate (HWP)~\cite{1988PASP..100.1158J,1991PASP..103.1193P,1991ApJ...370..257L,2007ApJ...665...42J,JohnRuhl2008} is one of the most promising modulation techniques to separate the CMB B-mode signal from the large unpolarized atmosphere for instruments with large optical throughput and large numbers of pixels.~\cite{2007ApJ...665...42J,2010SPIE.7741E..37R}

Here we report results from using a HWP on the Atacama B-Mode Search (ABS) instrument to modulate
the incident polarization at a rate well above the $1/f$ knee of the
atmospheric emission. To the best of our knowledge, ABS is the first experiment to use a rapidly-rotating HWP on a
ground-based CMB experiment. Using a demodulation procedure, we find
that the polarized signal band is free from atmospheric contamination
at the level of 0.1\% of that in the raw intensity signal.
   This allows the noise in the demodulated TOD to integrate
   down as $1/\sqrt{t}$ over the
  time scales of 1000 seconds and demonstrates the ability of
  ground-based CMB experiments to recover CMB polarization over large
  angular scales.
 In the constant-elevation scans done by ABS, this facilitates recovery of modes perpendicular to the scan direction with wavelengths up to 2--4$^{\circ}$, corresponding to multipoles $\ell >$ 40--90, even in the presence of strong atmospheric emission. This multipole range is crucial for cosmology as the inflationary gravitational-wave signal is expected to peak at $\ell \sim 100$.  

The technique of fast modulation also makes it unnecessary
to difference the timestreams of paired detectors sensitive to two orthogonal linear polarizations.
Without fast modulation, a ground-based experiment using bolometric polarimeters usually achieves 
some rejection of atmospheric fluctuations by pair differencing in the
analysis pipeline. This differencing requires that both detectors in a
pair be functional, with well-understood and stable responsivity ratios,
reducing detector array efficiency and overall sensitivity. For ABS, the two detectors in a pair can operate independently.

The paper is organized as follows. Section~\ref{sec:instrument} provides
a brief description of the ABS instrument. Section~\ref{sec:hwp_sync}
describes the modulation signal in the detector timestreams. In Section~\ref{sec:demod},
we review the demodulation technique applied to the data to isolate celestial polarization from the unpolarized background and sources of instrument polarization. Section \ref{sec:data_quality} evaluates the quality of the demodulated TODs. 

\section{The ABS Instrument}
\label{sec:instrument}
The Atacama B-mode Search (ABS) is a 145 GHz receiver
consisting of 240 feedhorn-coupled polarimeters with 480 transition-edge
sensors (TESes) that operate at a base temperature of 300 mK. The target
bandpass for ABS is 127--160 GHz; however, approximately half of the
array has a shifted bandpass attributed to unplanned changes in the
index of refraction of the silicon-dioxide microstrip dielectric on the
detector wafers. These detectors have a bandpass of 140--170 GHz with a
broader cutoff at the high-frequency side, making those detectors more
susceptible to contamination from the water line of the atmosphere at
183\,GHz.

 Cryogenic 60-cm primary and secondary reflectors in a crossed-Dragone configuration couple the focal plane to the sky with 35$^{\prime}$ full width half maximum (FWHM) beams and a 22$^{\circ}$ field of view. 
A 25-cm diameter aperture stop at 4\,K  terminates beam spill at a
stable and cold surface. ABS is located next to the Atacama Cosmology
Telescope (ACT)\cite{2007ApOpt..46.3444F,2011ApJS..194...41S} at an altitude of 5190\,m, on Cerro Toco in the Atacama Desert of northern Chile. 

ABS is unique among current and planned CMB polarization experiments in
modulating the incoming polarization using an ambient-temperature,
continuously-rotating HWP.~\cite{Note1, 2007PhDT........22L}
The HWP is made of single-crystal, $\alpha$-cut sapphire 330 mm in diameter and 3.15 mm
thick. It is anti-reflection (AR) coated with 305 $\mu$m of Rogers
RT/Duroid, a glass-reinforced, ceramic-loaded polytetrafluoroethylene (PTFE) laminate with a refractive index of 1.71. The HWP emission is estimated from literature values\cite{1994IJIMW..15..339P, Note2} to be 2.3\,K and 3.2\,K for the extraordinary and ordinary axes, respectively. This level of emission would degrade system sensitivity by approximately 6\%. In-field characterization of loading is on-going, with warm optics estimated to add 8\,K of loading. This is an upper limit on HWP loading, as we have reason to believe that some excess loading comes from sources other than the HWP, including diffraction onto warm elements or unexpected emission from the HWP anti-reflection coating. We are working to reduce this excess warm loading.

An air-bearing system allows the HWP for ABS to rotate at a stable frequency. 
Porous graphite pads\cite{Note3} are placed around an aluminum rotor at three points on its circumference.
Compressed air is forced through the graphite to float the rotor with almost no friction. An incremental encoder
disc with an index to mark the zero point is used to read out the HWP angle with $2.4^{\prime}$ resolution.

\begin{figure}[tbp]
 \begin{center}
  \includegraphics[width=0.46\textwidth]{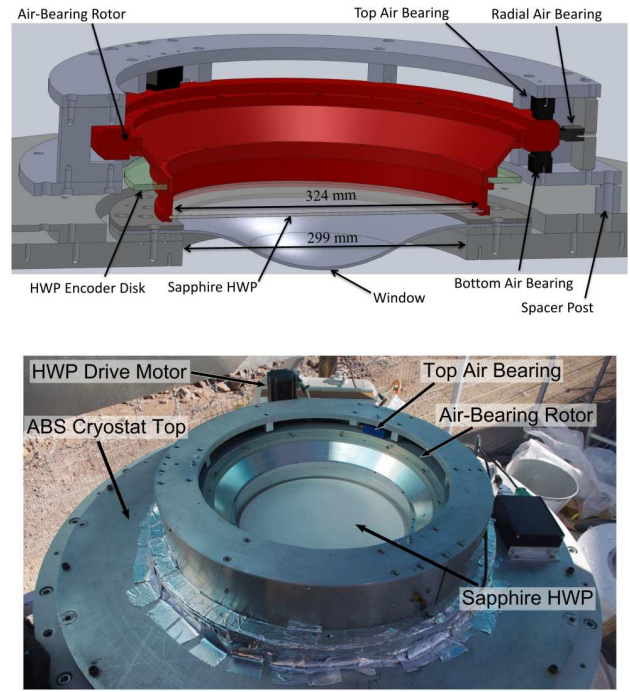}
  \caption{ \label{fig:hwp_photo} %
  Top: Cross-section drawing of the ABS HWP and air-bearing system showing
  the 3.2\,mm thick ultra-high molecular weight polyethylene (UHMWPE) vacuum window, sapphire HWP mounted in its rotor, air bearings, encoder disc, and the overall HWP support.
  Bottom: Photograph of the HWP installed on the ABS cryostat at the Chilean site.  
  The white-colored surface of the HWP is the AR-coating material.}
 \end{center}
\end{figure}

The HWP is placed at the entrance aperture of the telescope
directly above the vacuum window and is the first optical element in the
path of light from the sky to the focal plane, allowing for a
clear separation of sky polarization and instrument polarization.
The location is also designed to be the beam waist of the ensemble of the
detectors so that the detectors share the same HWP surface and some of
the systematics originating from the HWP can be removed as a common mode among detectors.
Figure~\ref{fig:hwp_photo} shows a drawing of the ABS HWP system and
the system in its installed configuration.
Further details on the ABS instrument can be found elsewhere~\cite{SuzanneAbs2008,2009AIPC.1185..494E,EssingerHilemanPrincetonThesis,JohnAppelPrincetonThesis,KaterinaVisnjicPrincetonThesis,Sara2013}.

\section{Modulation by HWP: Signals and Systematics}
\label{sec:hwp_sync}
\begin{figure}[tbp]
 \begin{center}
  \includegraphics[width=0.37\textwidth, trim=0pt -20pt 0pt 0pt ,clip]{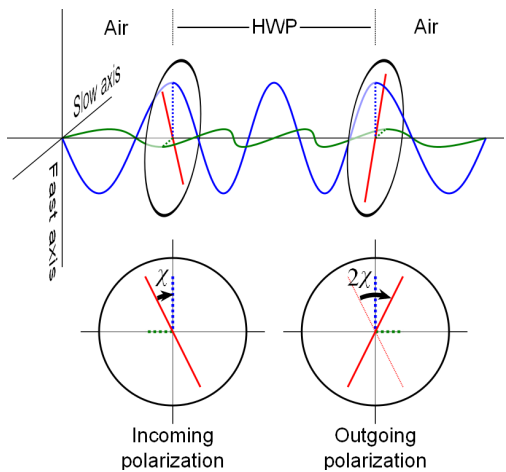}
  \caption{ \label{fig:hwp_prop} %Caption 
Propagation of a wave through the HWP. Incoming linear polarization is decomposed into two orthogonal linear polarizations along the crystal axes of the sapphire (fast and slow axes). These two waves travel at different speeds. The sapphire thickness is chosen to produce a 180$^{\circ}$ phase shift between these two waves, which reflects the incoming polarization about the fast axis of the crystal. 
  As a result, the polarization rotates by an angle of $2\chi$, where the angle between the fast axis and incoming polarization is $\chi$.
  The number of oscillations and relative wavelengths for the two waves within the sapphire as shown are merely illustrative. 
}
 \end{center}
\end{figure}

\begin{figure}[tbp]
% \begin{center}
	\centering
  \includegraphics[width=0.45\textwidth, trim=0pt -15pt 0pt 0pt ,clip]{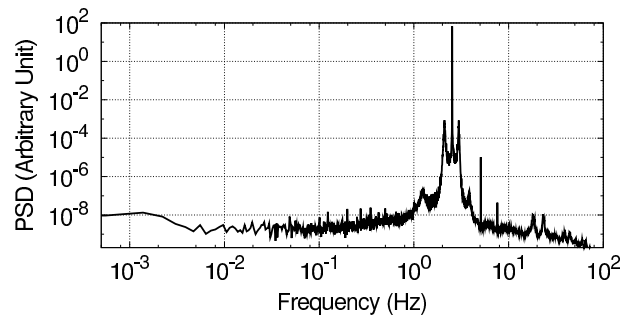}
  \caption{ \label{fig:hwp_encoder_fft} %
    The power spectral density (PSD) of the HWP encoder readout during a CMB observation of $\sim 1$\,hour duration.
    The sharpness of the main peak at $\sim 2.5$\,Hz emphasizes the stability of the rotation.
    The highly suppressed secondary peaks that are $\sim 0.3$\,Hz away from the main peak
    correspond to small and slow frequency modulation due to the servo cycles of the HWP rotation mechanism.
    Our analysis accounts for the exact HWP rotation including its frequency modulation as opposed to
    assuming a constant rotation frequency.
}
% \end{center}
\end{figure}

An ideal HWP rotates linear polarization by $2 \chi$, where $\chi$
is the angle between the incident polarization angle and the crystal
axis of the sapphire (Fig.~\ref{fig:hwp_prop}). 
ABS operates with $f_m \simeq 2.5\,\mathrm{Hz}$, where $f_m$ denotes the
HWP rotation frequency. This rotates the incident polarization at $2f_m\simeq 5\,\mathrm{Hz}$, which is detected in the bolometers at $4f_m \simeq 10\,\mathrm{Hz}$. 
The rotational frequency $f_m$ has a small modulation of order
$0.1\,\mathrm{Hz}$ (see Fig.~\ref{fig:hwp_encoder_fft}); however,
the treatment in this paper is general and does not assume a constant
modulation frequency.
With only the sky signal taken into account, the HWP-modulated signal,
$d_m(t)$, may be
represented in terms of the incoming Stokes parameters $I(t)$, $Q(t)$,
and $U(t)$, as well as the angle $\chi(t)$:
\begin{equation}
 d_{m} = I + \varepsilon \mathrm{Re} \{ (Q+\imath U ) m(\chi) \} \:.
 \label{equ:modulation_equation}
\end{equation}
Here, $\varepsilon$ is a polarization modulation efficiency factor,
which is close to unity for ABS \cite{KaterinaVisnjicPrincetonThesis, EssingerHilemanPrincetonThesis} and $m(\chi)$ is the modulation function:
\begin{equation}
 m(\chi) = \exp [ - \imath 4 \chi ] \:.
 \label{equ:modulation_equation2}
\end{equation}

\begin{figure}[tbp]
 \begin{center}
  \includegraphics[width=0.4\textwidth, trim=0pt -5pt 0pt 0pt ,clip]{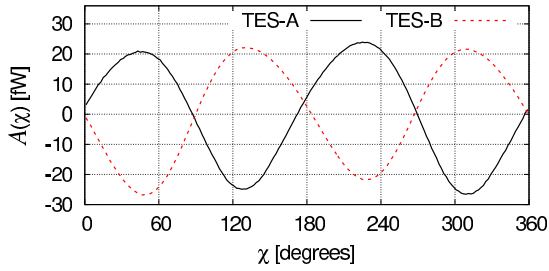}
  \caption{ \label{fig:A_of_chi_example} %
  Examples of $A(\chi)$ averaged over $\sim 1$\,hour of observation.  Two lines are shown, corresponding to two TESes (TES-A and TES-B) of a single pixel.  
  These curves are each clearly dominated by the $2f_m$ component, composed primarily of polarized emission from the HWP and differential transmission of unpolarized atmospheric emission.
  The two TESes are sensitive to orthogonal linear polarizations and the approximate sign flip between their $A(\chi)$ is the expected behavior.
  }
 \end{center}
\end{figure}

In practice, signals synchronous with the HWP rotation come from a
number of sources other than sky polarization.
Because the HWP is warm, the dominant HWP-synchronous signal comes from polarized emission due to differential emissivity of the sapphire along different crystal
axes, where the difference is roughly 0.3\,\%.\cite{1994IJIMW..15..339P} Differential transmission produces linear polarization from unpolarized sky emission and is the second most significant HWP-synchronous signal for ABS. Reflections of radiation from the receiver off the HWP and back to the detectors will be polarized in reflection. These effects produce a linear polarization fixed relative to the HWP optical axis that couples to the detectors at $2 f_{m}$. This signal can be modulated up to the $4 f_{m}$ signal band by small misalignments of the HWP axis from the air-bearing or encoder axes, as well as non-uniformities in the AR coating. At non-normal incidence, reflection can also produce a small polarization that can be modulated at $4 f_{m}$. These signals form part of the $A(\chi)$ in Eq. \ref{equ:modulated_dat_with_systematics} below. To the extent that the HWP motion is smooth and the AR coating uniform, these spurious sources of $4 f_{m}$ signal are expected to be small. As can be seen in Fig. \ref{fig:A_of_chi_example}, we find that this is indeed the case in the ABS data, where $A(\chi)$ does not have a significant $4 f_{m}$ component. 

The signal of interest, CMB polarization, occurs principally at
$4 f_{m}$ as shown in Eqs.~(\ref{equ:modulation_equation}) and (\ref{equ:modulation_equation2}). Any leakage of unpolarized power into the $4 f_{m}$ signal band is particularly detrimental because the unpolarized atmospheric signal and temperature anisotropies of the CMB are many orders of magnitude stronger than the signal of interest. Such $I \rightarrow Q/U$ leakage arises at non-normal incidence, as a small linear polarization is generated upon each reflection in the sapphire and its AR coating, which is then rotated by the sapphire. 

Given the 22$^{\circ}$ field of view of ABS, $I \rightarrow Q/U$ leakage
from off-axis incidence must be evaluated. A $4 \times 4$ transfer-matrix model that relates total electric and magnetic fields at the material boundaries was developed \cite{2010ApOpt..49.6313B,2013ApOpt..52..212E} to estimate possible $I \rightarrow Q/U$ leakage from the HWP. 
The model estimates the leakage to be less than $-30$\,dB.
The low $1/f$ knee in the demodulated data, see Section \ref{sec:data_quality}, is demonstration that this leakage is small, consistent with model expectations.  

Including these spurious modulation signals as $A(\chi)$, as well as white noise in the measurement, $\mathcal{N}_w$, Eq.~(\ref{equ:modulation_equation}) may be rewritten as
\begin{equation}
 d_{m} = I + \varepsilon \mathrm{Re} \{ (Q+\imath U ) m(\chi) \}
  + A(\chi) + \mathcal{N}_w\:.
  \label{equ:modulated_dat_with_systematics}
\end{equation}
We identify two contributions to $A(\chi)$:
\begin{equation}
 A(\chi) = A_0 (\chi) + \lambda(\chi) I \:,
\end{equation}

\noindent where $A_0 (\chi)$ is a component that is independent of sky intensity (e.g., the differential emissivity of the HWP) and the second term corresponds to the contribution of unpolarized sky signal to the HWP synchronous signal.
For later convenience, we decompose $A_0 (\chi)$ and $\lambda(\chi)$ as
\begin{equation}
 \begin{split}
 A_0 (\chi) & = \sum_n \left( A_0^{nC} \cos n \chi + A_0^{nS} \sin n \chi
		     \right)
  \:,
  \\
  \lambda (\chi) & = \sum_n \left( \lambda^{nC} \cos n \chi + \lambda^{nS} \sin n \chi
		     \right)
  \:.
  \\[10pt]
 \end{split}
\end{equation}

Fourier coefficients $\lambda^{4C}$ and $\lambda^{4S}$ correspond to
a $4f_m$ modulation in $d_m$ and thus the $I \rightarrow Q/U$ leakage.
The functional shape of $A(\chi)$ is predominantly constant in time
since $A_0(\chi)$ tends to be stable, $\lambda(\chi)$ is small,
and the fluctuation of $I$ is small compared to the absolute value
of $I$.  Figure~\ref{fig:A_of_chi_example} shows an example of the constant $A(\chi)$ components for the two orthogonal detectors in an ABS polarimetric detector.

Variation of the function $A(\chi)$ over long time scales occurs due to fluctuations in unpolarized atmospheric emission as well as the time variation of $A_0(\chi)$ caused by temperature drifts of the HWP or changes in detector responsivity. Such a variation of $A(\chi)$ leads to low frequency, so-called ``$1/f$,'' noise in the demodulated timestream.

\section{Demodulation Technique}
\label{sec:demod}

\begin{figure}[tbp]
 \begin{center}
  \includegraphics[width=0.48\textwidth, trim=0pt -10pt 0pt 0pt ,clip]{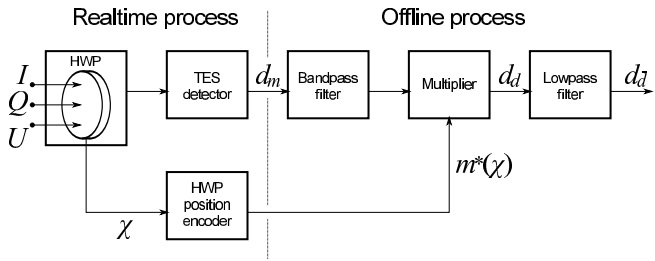}
  \caption{ \label{fig:hwp_demodulation_block_diagram} %
  A diagram of the data processing
  of a continuously rotating HWP modulation and demodulation.
  }
 \end{center}
\end{figure}

We will now use a demodulation technique to demonstrate the
stability of the ABS instrument in polarization, as well as the low
leakage of unpolarized atmospheric intensity into the demodulated
timestreams. The (unfiltered) complex demodulated timestream is created by multiplying the modulated data [Eq.~(\ref{equ:modulated_dat_with_systematics})] by the complex
conjugate of $m(\chi)$, the modulation function:

\begin{widetext}
\begin{equation}
 \begin{split}
 d_{d}  =  m^{\ast} d_{m} =& \quad \exp(\imath 4 \chi) \left[ I + \varepsilon Q \cos 4 \chi + \varepsilon U \sin 4 \chi  + A(\chi) + \mathcal{N}_w \right] \\
  %  &= \frac{1}{2} \left( Q + \imath U \right) + I \left( \cos 4 \chi + U \sin 4 \chi \right) + \frac{1}{2} \left( Q - \imath U \right) \left( \cos 8 \chi + \imath \sin 8 \chi \right) \\
   = & \quad \frac{\varepsilon}{2} \left( Q + \imath U \right) + \mathcal{N}_w^\mathrm{re} + \imath \mathcal{N}_w^\mathrm{Im} + I \exp[\imath 4 \chi] + \frac{\varepsilon}{2} \left( Q - \imath U \right) \exp[\imath 8 \chi] \quad \\
   & \quad +  \frac{1}{2} \sum_{n \geq 1} \left\{ \left(A^{\prime n C} - \imath A^{\prime n S} \right) \exp{[\imath (n+4) \chi]} + \left(A^{\prime n C} + \imath A^{\prime n S} \right) \exp{[- \imath (n-4) \chi]} \right\}
 \end{split} \:,
 \label{equ:master_equation_of_demod_signal_before_lowpass}
\end{equation}
\end{widetext}

\noindent where the complex coefficients ${A'}^{nC}$ and ${A'}^{nS}$ are linear
combinations of $A_0^{n C(S)}$ and
 $\lambda^{n C(S)} I$,
and $\mathcal{N}_w^\mathrm{Re}$ and
$\mathcal{N}_w^\mathrm{Im}$
 are the real and imaginary
parts of $m^* \mathcal{N}_w$, respectively.
Under our assumption of white noise for $\mathcal{N}_w$,
$\mathcal{N}_w^\mathrm{Re}$ and $\mathcal{N}_w^\mathrm{Im}$ are also white noise and their noise power satisfy:
\begin{equation}
 \left< \left| \mathcal{N}_w^\mathrm{Re} \right|^2 \right>
  = \left< \left| \mathcal{N}_w^\mathrm{Im} \right|^2 \right>
  = \frac{1}{2} \left< \left| \mathcal{N}_w \right|^2 \right>
  \:.
  \label{equ:noise_power_before_after_demod}
\end{equation}
Applying a lowpass filter to $d_d$ eliminates the terms that still depend on
$\exp[\imath 4 \chi]$ and $\exp[\imath 8 \chi]$ as well as all
$A^{\prime}$ terms other than the $n=4$ component; this leads to the
final demodulated timestream, denoted $d_{\bar{d}}$:
\begin{equation}
 \begin{split}
  d_{\bar{d}}
  = & \frac{1}{2} \left( \varepsilon Q + A_0^{4C} + \lambda^{4C} I  \right)+ \mathcal{N}_w^\mathrm{Re}
  \\
  &+ \frac{\imath}{2} \left( \varepsilon U + A_0^{4S} + \lambda^{4S} I  \right) + \imath \mathcal{N}_w^\mathrm{Im} \:.
 \end{split}
  \label{equ:demod_d_d_after_lowpass}
\end{equation}
For ABS's half-degree beam size and its scan speeds ($0.3$--$0.7^\circ/\mathrm{s}$ on the sky), most of the $Q$ and $U$ signals
from the CMB in the demodulated time stream are in the frequency range of $f \lesssim 1$\,Hz and the
passband of the lowpass filter  is set to $f \lesssim 2$\,Hz. The procedure is shown in Fig.~\ref{fig:hwp_demodulation_block_diagram}. Although it is not mentioned above, we apply a bandpass filter to $d_m$ before multiplying it by $m^*$ as indicated in the figure.
Note that in the ideal case of constant $f_m$, this filter is obviated
by the lowpass filter on $d_d$.  However, since for ABS $f_m$ varies slightly as shown in Fig.~\ref{fig:hwp_encoder_fft}, the bandpass filter effectively suppresses contamination that is localized in the frequency domain of $d_m$ (e.g., electric or magnetic pickup at certain frequency or the atmospheric $1/f$ noise).
Without the bandpass filter, the variation of $f_m$ smears this contamination in the frequency space of $d_d$.

In Eq.~(\ref{equ:demod_d_d_after_lowpass}), 
the real and imaginary parts of $d_{\bar{d}}$ correspond to $Q$ and $U$ polarizations with spurious components and noise. Time variations of $A_0^{4C(S)}$ and $\lambda^{4C(S)}I$ lead to long time-scale fluctuation, or $1/f$ noise, in the measurement.
However, since we measure $Q$ and $U$ simultaneously, there is no loss
in sensitivity due to the HWP modulation.  The sensitivity to the polarized component is the same as to the total intensity.
We note that at no point in the analysis do we take the difference
between the pair of orthogonally oriented detectors.

\section{Quality of Demodulated Timestream}
\label{sec:data_quality}
\begin{figure}[tbp]
 \begin{center}
  \includegraphics[width=0.40\textwidth, trim=0pt -20pt 0pt 0pt ,clip]{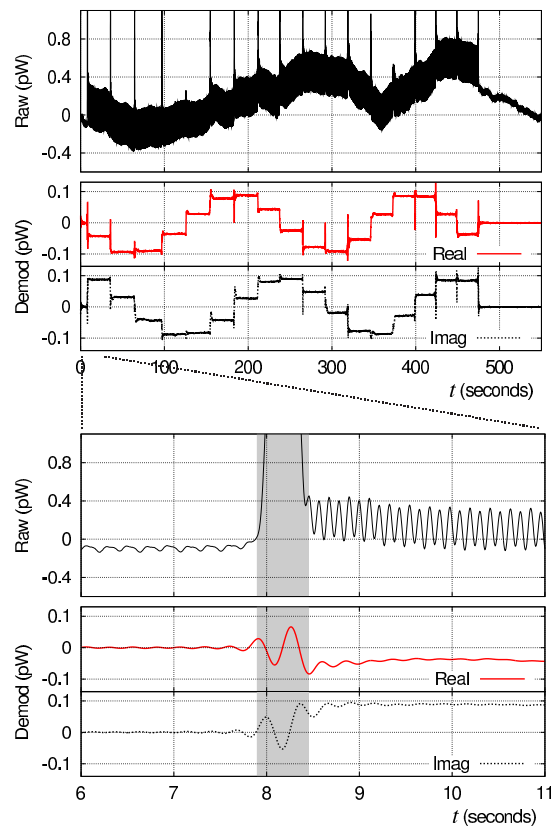}
  \caption{ \label{fig:wiregrid_demod} %
  Raw and demodulated timestreams from a single
  polarization-sensitive TES
  during a calibration session using a sparse wiregrid polarizer.
  The responsivity of the
  detectors is typically $\sim 100\,\mathrm{aW/mK}$.
  The top panel shows the entire calibration session.  It consists of 17
  segments and the wiregrid remains stationary for
  about 30\,seconds during each segment.
  The wiregrid is rotated by
  $\simeq 22.5\degs$ between the segments and thus it revolves by
  $360\degs$ during
  the entire session, where the first and last segments correspond to
  the same wiregrid position.
  As expected, the real and imaginary parts of the demodulated timestream
  show two periods of oscillation and have a phase difference of
  $45\degs$ with respect to each
  other (i.e., their phases are shifted by two segments). 
  The baseline  drift seen  in the raw data is  rejected by demodulation.
  The bottom panel zooms in on the initial part of the calibration.
  The wiregrid was placed during the period of $t \simeq
  7.9$--$8.5$\,sec
  (shaded area in the panel).
  In the raw timestream, one can see a small modulation even while the
  wiregrid was
  absent ($t<7.9$\,sec), which comes from the $A(\chi)$ 
  part in Eq.~(\ref{equ:modulated_dat_with_systematics}) 
  and manifests mainly in the $2 f_m$ signal.  
  }
 \end{center}
\end{figure}

We now investigate the quality of the demodulated timestreams during select calibration runs and normal CMB observations. Figure~\ref{fig:wiregrid_demod} shows the raw and demodulated timestreams of a calibration session using a sparse wiregrid similar to that developed for the Q/U Imaging ExperimenT (QUIET).~\cite{2012JLTP..167..936T} The wiregrid is comprised of thin, reflective wires placed at intervals of one inch.  During a calibration session, we placed the wiregrid on the plane perpendicular to the line of sight, and rotated it discretely around the line of sight to vary the polarization angle of the radiation reflected by the grid.  As shown in Fig.~\ref{fig:wiregrid_demod}, the HWP modulates the signal in the raw timestream, and the demodulation procedure correctly reconstructs the injected polarization signal. The figure also shows that the baseline drift seen in the raw timestream is highly suppressed in the demodulated timestream.

We apply the same procedure to data taken during a CMB observation. During the observation, the telescope scans the sky periodically in azimuth, while the elevation axis is fixed. The azimuth scans typically have a frequency of $\sim 0.04\,\mathrm{Hz}$ and are repeated for 1--1.5\,hours; a set of repeated azimuth scans is denoted a ``constant-elevation scan'' (CES). The detectors are rebiased at the beginning of every CES.
In evaluating the data, we apply crude data selection criteria to reject
ill-behaved TESes (e.g., those that are not properly biased, those with too many glitches, etc.). We also reject TESes that have significantly low optical efficiency ($\sim 20\%$ of a typical TES or lower) since inclusion of those may overestimate the stability of the timestreams.
Typically $\sim 300$ TESes (out of $\sim 400$ functional TESes) pass these
criteria and are used for further evaluation.

\begin{figure}[tbp]
 \begin{center}
  \includegraphics[width=0.45\textwidth, trim=0pt -10pt 0pt 0pt ,clip]{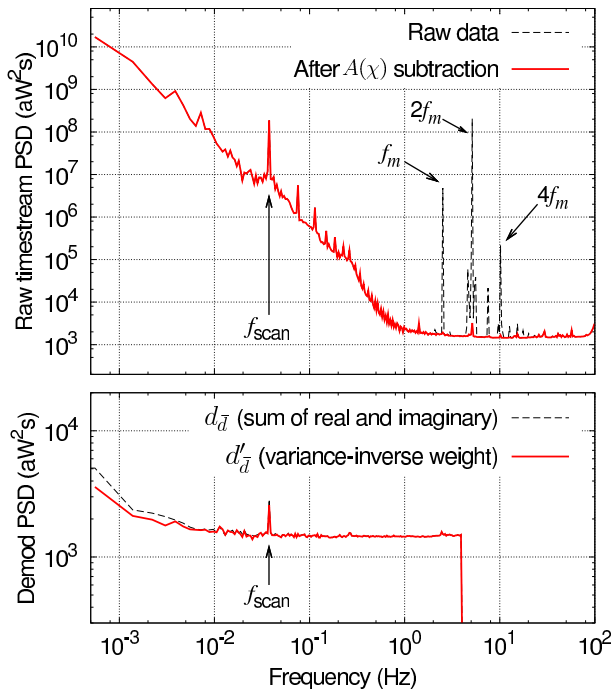}
  \caption{\label{fig:spectrum_example} %
  The power spectra of the TES timestreams
  before and after the demodulation.
  The data are from the same CMB-observing CES as
  Fig.~\ref{fig:hwp_encoder_fft}.
  Each spectrum is the inverse-variance weighted average over $\sim 300$
  TESes.
  The top panel shows the power spectra of the timestreams before demodulation.
  The dashed and solid lines correspond to the timestreams before and 
  after the subtraction of $A(\chi)$ that is constant during this CES, respectively.
  The bottom panel shows the power spectra of the demodulated
  timestream.
  For the dashed line, the spectrum of each TES
  is a sum of the power spectra of real and imaginary parts of the
  demodulated timestream $d_{\bar{d}}$.
  For the solid line, the spectrum of each TES is obtained as a
  inverse-variance weighted sum of the spectra of the real and imaginary parts
  of $d'_{\bar{d}} \equiv e^{\imath \phi_0} d_{\bar{d}}$,
  which is defined in the text.
  The typical responsivity of a TES is $\sim 100\,\mathrm{aW/mK}$.
  The scan frequency ($f_\mathrm{scan}$) as well as the harmonics of the
  modulation frequency ($f_m$) are indicated by arrows.
  }
 \end{center}
\end{figure}

Figure~\ref{fig:spectrum_example} shows the power spectra averaged for
$\sim 310$ TES timestreams from a CES performed on Oct. 6th, 2012, when
the precipitable water vapor (PWV) was $\sim 0.7$\,mm~\cite{Note3}.
As can be seen in the demodulated power spectrum (bottom), 
the ABS polarimeters are white-noise dominated above a few mHz.

The average among TESes is calculated by inverse-variance weighting the
power spectrum of each TES in each frequency bin so that it reflects the effective noise level relevant in the science analysis. 
The top panel of Fig.~\ref{fig:spectrum_example} shows the power spectra before the demodulation.
  We show both the raw timestream spectrum and the spectrum after the subtraction of the $A(\chi)$ component that is constant over the CES.  Although large spikes are observed at the harmonics of the modulation frequency $f_m$, most of them correspond to the time-invariant $A(\chi)$ and disappear after the subtraction. Note that the subtraction leads to little loss of the CMB signal.~\cite{Note4}
The bottom panel shows the power spectra of the demodulated timestreams.
Here, there are two power spectra that correspond to two different methods
of obtaining the power spectrum of a single TES.
One is to simply Fourier transform each of the real and imaginary parts
of $d_{\bar{d}}$, and add them in power, which we will simply refer to as
``the spectrum of $d_{\bar{d}}$.''
  In the other method,
we first redefine the
demodulated timestream as $d'_{\bar{d}} \equiv e^{\imath \phi_0} d_{\bar{d}}$,
where $\phi_0$ is found by maximizing (minimizing) the low-frequency
noise of the real (imaginary) part of $d'_{\bar{d}}$,
and then averaging the power spectra of the real and imaginary
parts by inverse-variance weighting in each frequency bin
in the same manner as we do in averaging different TESes.
Note that the demodulated timestream from a single TES has two degrees
of freedom (the real and imaginary parts)
and these two correspond to
two orthogonal linear combinations of $Q$ and $U$
polarizations.
Thus, $d'_{\bar{d}}$ simply redefines the orthogonal linear
combinations such that they are the eigenmodes of the $1/f$ noise for a
single TES.
The phase $\phi_0$ is determined separately in each CES for each TES.
We typically find only one dominant mode of $1/f$ noise in the two degrees of freedom of a demodulated timestream, and the imaginary part of $d'_{\bar{d}}$ tends to exhibit extremely low $1/f$ noise. Thus, the second method of inverse-variance weighting $d'_{\bar{d}}$ evaluates the effective noise level better than the first method.
The phase $\phi_0$ for each TES tends to show
 large CES-by-CES scatter.
However, it tends to show a mild peak at a certain value and the peak
is correlated to the polarization angle to which the TES is
sensitive.  Thus, at least part of the source of the $1/f$ noise has
optical origin such as $I \rightarrow Q/U$ leakage.

As shown in the top panel of Fig.~\ref{fig:spectrum_example}, the low-frequency noise component in raw timestreams typically has $f_\mathrm{knee} \sim 1\,\mathrm{Hz}$ and a spectral index $\alpha \sim 2$.  The $\sim 1$\,mHz knee frequencies in the demodulated timestreams correspond to $\sim 10^6$ reduction in noise power (or $\sim 10^3$ reduction in amplitude) and demonstrate a leakage from total power to polarization of less than $-30$\,dB.

\begin{figure}[tbp]
 \begin{center}
  \includegraphics[width=0.45\textwidth, trim=0pt -20pt 0pt 0pt ,clip]{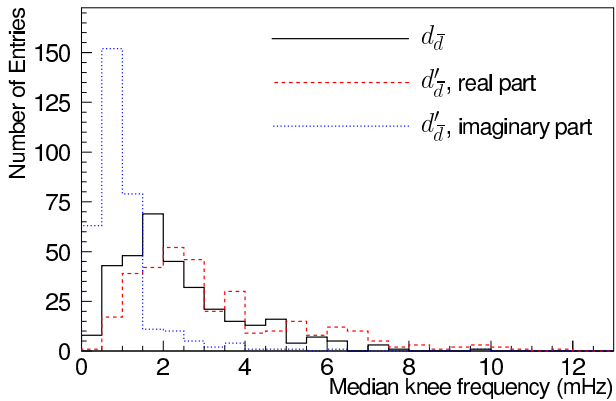}
  \caption{ \label{fig:fknee_distr} %
  Distribution of the knee
  frequencies of the demodulated timestream
  for $\sim 330$ detectors.
  Each entry corresponds to a single detector;
  for each detector, the median knee frequency is calculated
  by evaluating science data taken during a one month period in
  October and November 2012, when the median PWV was $\sim 0.7$\,mm.
  The three distributions correspond to 
  the knee frequencies of the following three power spectra:
  the sum of the real and imaginary parts of $d_{\bar{d}}$
  (solid line),
  the real part of $d'_{\bar{d}}$ (dotted line),
  and the imaginary part of $d'_{\bar{d}}$ (dashed line).
  The imaginary part of $d'_{\bar{d}}$
  shows extremely low knee frequencies, implying that there is
  only one dominating $1/f$ noise mode for the two degrees of freedom of
  the demodulated timestream.
  }
 \end{center}
\end{figure}

Figure~\ref{fig:fknee_distr} shows the distribution of the knee
 frequencies of the demodulated timestreams for $\sim 330$ detectors.
The knee frequencies, characterizing the levels of low-frequency noise
 excess,
 are determined by fitting the following model to the power spectrum:
\begin{equation}
 P(f) = P_0 \left[ 1 + \left( \frac{f_\mathrm{knee}}{f} \right)^\alpha \right],
\end{equation}
where free parameters of the fit are the high-frequency noise power $P_0$,
the spectral index $\alpha$, and the knee frequency $f_\mathrm{knee}$.
In the fit, the region around the scan frequency $f_\mathrm{scan}$ is
excluded.
Three knee frequencies are determined for each detector by
fitting the spectrum of $d_{\bar{d}}$ as well as the spectrum for each of the
real and imaginary parts of $d'_{\bar{d}}$.
 We evaluate the primary science data of about one month period. 
 For each detector, the median knee frequency over this period is calculated
 for each of the three spectra.
 Shown are the distributions of these median knee frequencies per
 detector. The medians among all the detectors are 2.0\,mHz,
 2.6\,mHz and 0.8\,mHz for
 the spectra of $d_{\bar{d}}$,
 the real part of $d'_{\bar{d}}$,
 and the imaginary part of $d'_{\bar{d}}$,
 respectively.

Since the $f_{\mathrm{knee}}$ is so low that its timescale is close to the
length of the timestream itself, small biases are introduced in the estimates above.  These biases are evaluated
through Monte-Carlo simulation studies and all of the results presented
above are corrected for the bias.  The corrections are small, corresponding to $\sim 0.1$--$0.2$\,mHz in the estimation of $f_{\mathrm{knee}}$.

\section{Conclusion}
We have demonstrated the viability of the rapid rotation of CMB polarization using an ambient-temperature HWP. A demodulation technique based on mixing a complex modulation function with the time-ordered detector data is used to evaluate the quality of the science data from ABS.
This analysis shows that ABS has achieved mHz stability in its polarization measurement and leakage from total power to polarization lower than $-30$\,dB. 

\section*{Acknowledgments}
Work at Princeton University is supported by the U.S. National Science
 Foundation through
awards PHY-0355328 and PHY-085587, the U.S. National Aeronautics and
 Space Administration (NASA) through award NNX08AE03G,
the Wilkinson Fund, and the Mishrahi Gift.  
Work at NIST is supported by the NIST Innovations in
Measurement Science program. 
ABS operates in the Parque Astron\'{o}mico
Atacama in northern Chile under the auspices of the
Comisi\'{o}n Nacional de Investigaci\'{o}n Cient\'{i}fica y
Tecnol\'{o}gica de Chile (CONICYT).
PWV measurements were provided by the Atacama
Pathfinder Experiment (APEX).
Some of the analyses were performed on the GPC supercomputer at the SciNet HPC
Consortium. SciNet is funded by the Canada Foundation of Innovation under the auspices of Compute
Canada, the Government of Ontario, the Ontario Research Fund --
Research Excellence; and the University of Toronto.
We would like to acknowledge the following people for
their assistance in the instrument design, construction, operation, and
data analysis: G.~Atkinson, J.~Beall, F.~Beroz, S.~M.~Cho, S.~Choi, B.~Dix, T.~Evans, J.~Fowler, M.~Halpern, B.~Harrop,  M.~Hasselfield, S.~P.~Ho, J.~Hubmayr, T.~Marriage,  J.~McMahon, M.~Niemack, S.~Pufu, M.~Uehara, K.~W.~Yoon.  
T.~E.-H. was supported by a National Defense Science and Engineering Graduate (NDSEG) Fellowship, as well as a National Science Foundation Astronomy and Astrophysics Postdoctoral Fellowship.
A.~K. acknowledges the Dicke Fellowship.
S.~M.~S. is supported by a NASA Office of the Chief Technologist's Space Technology Research Fellowship.
L.~P.~P. acknowledges the NASA Earth and Space Sciences Fellowship.
S.~R is in receipt of a CONICYT PhD studentship.
S.~R received partial support from a CONICYT Anillo project (ACT No. 1122).

\end{document}